\documentclass[aps,prb,reprint,amsfonts,amsmath,amssymb,longbibliography]{revtex4-2}
\usepackage{hyperref}
\usepackage{graphicx}
\usepackage{bm}
\usepackage{newtxtext}
\usepackage[varg]{newtxmath}
\usepackage{lineno}
\usepackage[normalem]{ulem}

\DeclareMathOperator{\sgn}{sgn}
\DeclareMathOperator{\Tr}{Tr}
\DeclareMathOperator{\real}{Re}

\hypersetup{colorlinks=true,linkcolor=blue,citecolor=blue,urlcolor=blue}

\begin{document}

\title{Ultrafast reorientation of the N\'eel vector in antiferromagnetic Dirac semimetals}
\author{Atsushi Ono}
\author{Sumio Ishihara}
\thanks{Deceased.}
\affiliation{Department of Physics, Tohoku University, Sendai 980-8578, Japan}

\begin{abstract}
Antiferromagnets exhibit distinctive characteristics such as ultrafast dynamics and robustness against perturbative fields, thereby attracting considerable interest in fundamental physics and technological applications.
Recently, it was revealed that the N\'eel vector can be switched by a current-induced staggered (N\'eel) spin-orbit torque in antiferromagnets with the parity-time symmetry, and furthermore, a nonsymmorphic symmetry enables the control of Dirac fermions.
However, the real-time dynamics of the magnetic and electronic structures remain largely unexplored.
Here, we propose a theory of the ultrafast dynamics in antiferromagnetic Dirac semimetals and show that the N\'eel vector is rotated in the picosecond timescale by the terahertz-pulse-induced N\'eel spin-orbit torque and other torques originating from magnetic anisotropies.
This reorientation accompanies the modulation of the mass of Dirac fermions and can be observed in real time by the magneto-optical effects.
Our results provide a theoretical basis for emerging ultrafast antiferromagnetic spintronics combined with the topological aspects of materials.
\end{abstract}

\maketitle

\clearpage
\section*{Introduction} \label{sec:intro}
The control and detection of antiferromagnetic (AFM) orders have been a challenging problem because of the absence of net magnetizations.
Since the energy scale of antiferromagnets lies in the terahertz range, which is beyond the scope of conventional electronics, optical pulses are suitable for controlling AFM orders~\cite{Nemec2018,Song2018,Kampfrath2013} through magnon excitations~\cite{Kampfrath2011,Nishitani2013,Baierl2016b,Li2020}, inverse magneto-optical effects~\cite{Kimel2005,Kimel2009,Satoh2015}, magnetic anisotropy~\cite{Duong2004,Kimel2004,Baierl2016,Schlauderer2019}, photoinduced phase transitions~\cite{Ju2004,Ehrke2011,Li2013,Ishihara2019}, and so forth~\cite{Takayoshi2014b,Mentink2015,Ono2017,Werner2019,Takasan2018}.
For detecting AFM orders, the magneto-optical effects are frequently utilized for collinear and canted antiferromagnets~\cite{Cheong2020,Saidl2017,Tzschaschel2019,Higo2018,Zheng2018b,Xu2019,Schreiber2020}.
The X-ray magnetic linear dichroism with photoemission electron microscopy has succeeded in imaging the AFM domain structure~\cite{Stohr1999,Scholl2000,Ohldag2001,Grzybowski2017}.
While these optical and X-ray measurements have the potential for time-resolved observations of magnetic dynamics~\cite{Nemec2018,Wen2019,Choe2004,Litzius2017,Dean2016}, their application to the AFM orders remains difficult and under development.

Several years ago, another mechanism that modulates AFM orders via electric currents was proposed and demonstrated~\cite{Zelezny2014,Wadley2016}.
In antiferromagnets with a combined space-inversion and time-reversal symmetry, electric currents induce a staggered spin density, and thus, a staggered torque termed the N\'eel spin-orbit torque (NSOT).
The NSOT can efficiently switch the staggered magnetization, i.e., the N\'eel vector, thereby yielding significant progress in AFM spintronics~\cite{Jungwirth2016,Baltz2018,Manchon2019,Amin2020}.
Recently, it was found that a nonsymmorphic crystalline symmetry in addition to the parity-time symmetry preserves the crossing of doubly degenerate bands; this implies that Dirac fermions are controlled by the direction of the N\'eel vector~\cite{Smejkal2017}.
The discovery of this close relationship between the magnetic and electronic structures opened a research field called topological AFM spintronics~\cite{Smejkal2018,Tsai2020}.
Some experiments have shown that not only electric current pulses but optical pulses can control the AFM order~\cite{Olejnik2018,Kaspar2020}; these studies performed transport measurements such as the anisotropic magnetoresistance and the planar Hall effect to observe the magnetic structure~\cite{Amin2020}.
Meanwhile, the topological electronic structure has received less attention, and therefore, the real-time dynamics of the N\'eel vector and the Dirac fermions induced by the NSOT are still unclear.

From a theoretical viewpoint, two approaches have been established to study the real-time dynamics of magnets.
One is based on microscopic models such as the Hubbard model and Kondo lattice model~\cite{Ishihara2019}, which allows considering electron correlation and quantum effects; however, the cluster size is severely limited and approximations are needed to deal with the symmetry-broken ordered states.
The other approach is to use the Landau--Lifshitz--Gilbert (LLG) equation of the classical vectors of magnetic moments, which has been widely adopted in spintronics~\cite{Maekawa2017,Blachowicz2019,Abert2019,Hals2011,Roy2016,Lopez-Dominguez2019}.
This equation enables the simulation of large-scale magnetic structures, while the electronic degree of freedom is integrated out.
Therefore, a unified framework is required to reveal the relationship between the magnetic and electronic structures.

In this work, we propose a theory based on a minimal two-dimensional model of AFM Dirac semimetals to investigate the real-time magnetic and electronic dynamics.
We show that the N\'eel vector is rotated by optically-induced NSOT and other torques originating from magnetic anisotropies.
This reorientation accompanies the modulation of the mass of Dirac fermions.
Furthermore, the magneto-optical effect is found to be promising for the time-resolved measurement of the N\'eel vector.

\section*{Results} \label{sec:results}
\subsection*{Theoretical model}

\begin{figure}[t]\centering
\includegraphics[width=1\hsize]{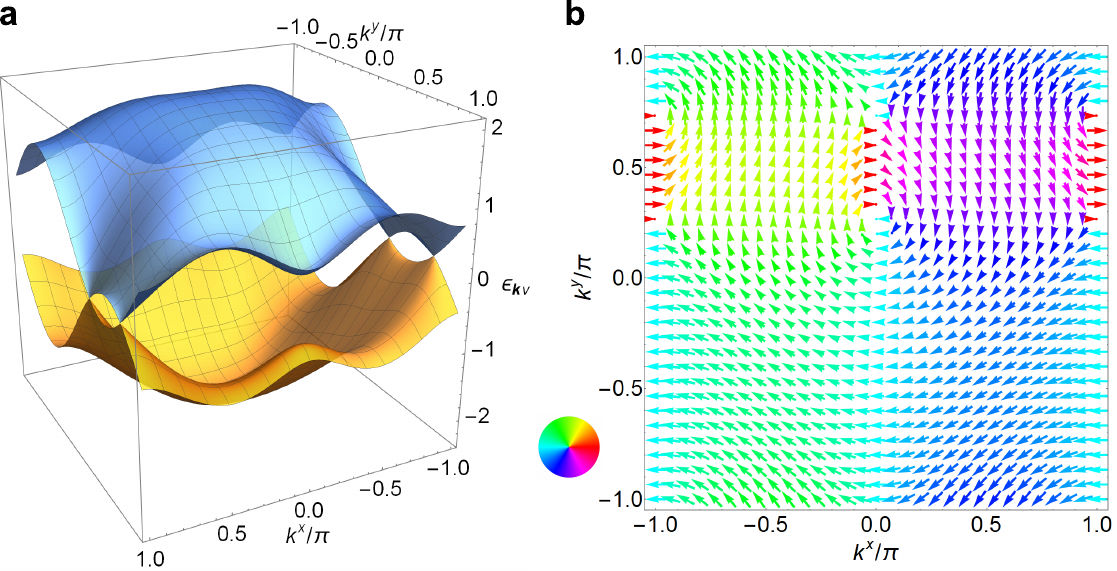}
\caption{Equilibrium properties of the model of AFM Dirac semimetals.
\textbf{a}~Energy band structure.
\textbf{b}~Staggered spin texture of the lower bands in the reciprocal space.
The localized spins are aligned antiferromagnetically along the $[100]$ direction, i.e., $\bm{n}=(1,0,0)$.
}
\label{fig:equil}
\end{figure}

We consider a minimal tight-binding model of AFM Dirac semimetals projected onto the two-dimensional square lattice.
The Hamiltonian is divided into three parts as $\mathcal{H}=\mathcal{H}_\mathrm{ele}+\mathcal{H}_\mathrm{exc}+\mathcal{H}_\mathrm{mag}$.
The first term
\begin{align}
\mathcal{H}_\mathrm{ele} &= -2h_1\tau^x \cos \frac{k^x}{2} \cos \frac{k^y}{2} -h_2(\cos k^x+\cos k^y) \notag \\
&\quad +\lambda\tau^z(\sigma^y\sin k^x-\sigma^x\sin k^y) \label{eq:hamiltonian_ele}
\end{align}
consists of the nearest-neighbor ($h_1$) and next-nearest-neighbor ($h_2$) hoppings, and the spin-orbit coupling ($\lambda$) of itinerant electrons with momentum $\bm{k}$.
Here, the spin and sublattice degrees of freedom of the itinerant electrons are described by the Pauli matrices $\bm{\sigma}$ and $\bm{\tau}$, respectively.
The localized magnetic moments on each sublattice denoted by $\bm{S}_\mathrm{A}$ and $\bm{S}_\mathrm{B}$ couple to the itinerant electrons through the exchange interaction
\begin{align}
\mathcal{H}_\mathrm{exc} &= J_\mathrm{exc} {\left[\frac{1+\tau^z}{2}\bm{\sigma}\cdot\bm{S}_\mathrm{A}+\frac{1-\tau^z}{2}\bm{\sigma}\cdot\bm{S}_\mathrm{B}\right]}. \label{eq:hamiltonian_exc}
\end{align}
Hereafter, we define the uniform and staggered magnetizations as $\bm{m}=(\bm{S}_\mathrm{A}+\bm{S}_\mathrm{B})/2$ and $\bm{n}=(\bm{S}_\mathrm{A}-\bm{S}_\mathrm{B})/2$, respectively.
The vector $\bm{n}$ is termed the N\'eel vector.
The exchange Hamiltonian is an extension of the model for $\bm{S}_\mathrm{A}=-\bm{S}_\mathrm{B}=\bm{n}$ proposed by \v{S}mejkal \textit{et al}.~\cite{Smejkal2017}.
While both the parity ($\mathcal{P}$) and time-reversal ($\mathcal{T}$) symmetries are broken because of the presence of the localized magnetic moments, the parity-time ($\mathcal{PT}$) symmetry is preserved when $\bm{m}=0$.
This guarantees the Kramers degeneracy of the energy bands in the whole Brillouin zone.
In addition to the $\mathcal{PT}$ symmetry, this model is invariant under a nonsymmorphic glide symmetry operation only when the N\'eel vector is directed along the $\langle 100\rangle$ directions, which causes the crossings of the doubly degenerate energy bands, i.e., the gapless Dirac points, to be protected at the Brillouin zone boundary~\cite{Smejkal2017}, as shown in Fig.~\ref{fig:equil}a.
The energy gap at the Dirac points depends on the value of $J_\mathrm{exc}$ and the direction of $\bm{n}(t)$.
The spin-orbit coupling and the AFM ordering of the localized moments lead to the spin-momentum locking of the itinerant electrons.
The expectation value of the staggered spin moments of the itinerant electrons in the lower bands is displayed in Fig.~\ref{fig:equil}b, which indicates that the net staggered moment $\langle \tau^z \sigma^x \rangle$ takes a negative value when $\bm{n}=(1,0,0)$ because of the AFM exchange interaction.

To stabilize the initial state in which the N\'eel vector is aligned with the $[100]$ direction, we introduce magnetic anisotropy energy
\begin{align}
\mathcal{H}_\mathrm{mag} &= \sum_{\gamma=\mathrm{A},\mathrm{B}} {\left[ K_z (S_\gamma^z)^2 + K_{xy} (S_\gamma^x S_\gamma^y)^2 \right]}, \label{eq:hamiltonian_mag}
\end{align}
where $K_{z}$ and $K_{xy}$ represent coefficients of the easy-plane and biaxial anisotropies, respectively.
Here, other magnetic interactions such as the exchange interaction between the localized moments, which can be implemented in a straightforward manner, are omitted because the AFM state is stabilized by the indirect exchange mediated by the itinerant electrons at half filling.

The time-dependent electric field $\bm{F}(t)$ parallel to the two-dimensional plane is incorporated in the model via the Peierls substitution.
The time evolutions of the itinerant electrons and the localized magnetic moments are described by the von Neumann and LLG equations, respectively; these equations are solved simultaneously by the fourth-order Runge--Kutta method.
The parameter values and units are presented in Methods.

\subsection*{Application of a constant electric field}

\begin{figure*}[t]\centering
\includegraphics[width=1\hsize]{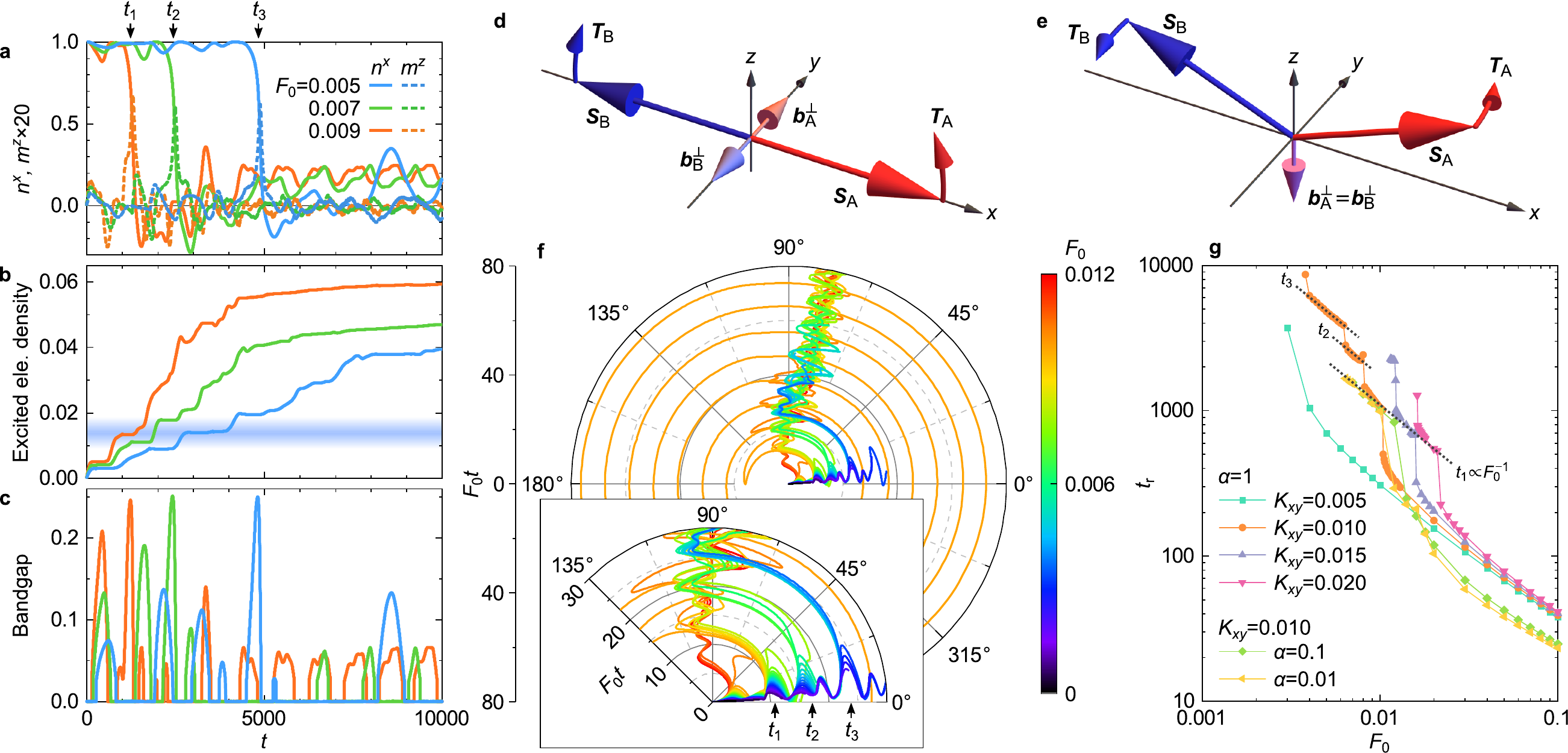}
\caption{Real-time dynamics induced by the constant electric field.
\textbf{a}--\textbf{c} The time evolution of (\textbf{a}) the N\'eel vector and uniform magnetization, (\textbf{b}) excited electron density, and (\textbf{c}) energy bandgap, as functions of time $t$ in units of $\hbar/h_1 = 0.66\ \mathrm{fs}$.
\textbf{d}~Sketch of NSOTs $\bm{T}_\mathrm{A}$ and $\bm{T}_\mathrm{B}$ acting on sublattice moments $\bm{S}_\mathrm{A}$ and $\bm{S}_\mathrm{B}$, respectively.
Arrows of $\bm{b}_\mathrm{A}^\perp$ and $\bm{b}_\mathrm{B}^\perp$ represent effective fields perpendicular to $\bm{S}_\mathrm{A}$ and $\bm{S}_\mathrm{B}$, respectively.
Uniform magnetization appears in the $[001]$ direction because of NSOT, and it induces another effective field attributed to the easy-plane anisotropy that rotates the N\'eel vector as shown in \textbf{e} with the $z$ components exaggerated.
\textbf{f}~Polar plots of the in-plane angle of the N\'eel vector as a function of the dimensionless time $F_0 t$ for $F_0=0.0004$--$0.012$.
The inset of \textbf{f} is the enlarged view from $F_0t=0$ to $30$.
\textbf{g}~Time $t_\mathrm{r}$ for the N\'eel vector to rotate towards the $[010]$ (or $[0\bar{1}0]$) direction as a function of $F_0$.
The electric field is applied in the $[110]$ direction in \textbf{a}--\textbf{g}.
}
\label{fig:dc}
\end{figure*}

First, we discuss real-time dynamics induced by a constant electric field that is switched on at time $t=0$ and applied in the $[110]$ direction, i.e., $\bm{F}(t) = F_0 \varTheta(t) (\cos\pi/4, \sin\pi/4, 0)$ with $\varTheta$ being the step function.
Figure~\ref{fig:dc}a shows the real-time profiles of the in-plane $x$-component of the N\'eel vector and the out-of-plane $z$-component of the uniform magnetization.
Figure~\ref{fig:dc}b displays the excited electron density defined by the number density of electrons that occupy single-particle energy levels above the chemical potential $\mu$ of the initial state.
The energy bandgap shown in Fig.~\ref{fig:dc}c is defined by the difference between the single-particle energies of the lowest level above $\mu$ and the highest level below $\mu$; this difference is equal to the direct gap when there is the particle-hole symmetry with $h_2=0$.
As time evolves, $n^x$ abruptly decreases towards zero at time $t=t_1$, $t_2$, or $t_3$, which is indicated by arrows in Fig.~\ref{fig:dc}a and is hereafter termed a reorientation time $t_\mathrm{r}$.
Meanwhile, the $z$ component of the uniform magnetization appears with values of the order of $m^z \sim 0.03$; this means that the magnetic structure is slightly canted as illustrated in Fig.~\ref{fig:dc}e.
As the field magnitude increases from $F_0=0.005$ to $0.009$, the reorientation time $t_\mathrm{r}$ becomes short.
The excited electron density shown in Fig.~\ref{fig:dc}b increases as time evolves.
Figure~\ref{fig:dc}b shows some plateaus in which the excited electron density is constant in time and the energy bandgap opens simultaneously (Fig.~\ref{fig:dc}c).
Once the excited electron density becomes greater than approximately $0.01$ (indicated by the shade in Fig.~\ref{fig:dc}b), the N\'eel vector starts to rotate.
Note that the bandgap opens through the N\'eel-vector dynamics induced by the constant field, which is in contrast to the gap opening of the Floquet--Bloch bands under the high-frequency driving of electrons~\cite{Oka2009}.
These phenomena induced by the constant electric field occur in a timescale of the order of $t\sim 10000 \hbar/h_1$, which corresponds to $6.6\ \mathrm{ps}$ for $h_1=1\ \mathrm{eV}$.
Therefore, the N\'eel vector and the energy bandgap can be controlled by the external field in such a short timescale.

The mechanism of the reorientation of the N\'eel vector is understood as follows.
The external electric field induces the nonequilibrium staggered spin density of the itinerant electrons through the spin-orbit coupling, and it generates torques in the $[001]$ direction (see Fig.~\ref{fig:dc}d).
Then, the two sublattice magnetic moments are slightly canted towards the $[001]$ direction, which means $m^z> 0$.
The easy-plane anisotropy, whose energy per sublattice moment is estimated to be $K_z(m^z)^2$, induces another effective field $(0,0,-2K_zm^z)$ that rotates the localized moments in the anticlockwise direction as depicted in Fig.~\ref{fig:dc}e.
Here, we have shown the NSOT-driven reorientation of the N\'eel vector on the basis of the microscopic model in which the real-time dynamics of the itinerant electrons and the localized moments are explicitly considered.

Figure~\ref{fig:dc}f shows the polar plots of the in-plane angle of the N\'eel vector $\varphi$ as a function of time multiplied by the field magnitude for different values of $F_0$.
The N\'eel vector rotates towards the $[010]$ direction for sufficiently large field magnitudes $F_0 \gtrsim 0.004$.
For the weak magnitude $F_0\lesssim 0.004$, the trajectories of $\varphi(F_0t)$ are on a universal curve independent of $F_0$; this implies that the timescale of the dynamics is determined by the vector potential $\bm{A}\sim \bm{F} t$ rather than the electric field $\bm{F}$.
This is because the NSOT is induced by the electric current due to the diamagnetic response.
The reorientation is regarded as a deviation from the weak-field universal trajectory.
Even before the reorientation, the N\'eel vector slightly deviates from the $x$ axis because of the NSOT; once the excited electron density exceeds approximately $0.01$, the NSOT-induced staggered torque shown in Fig.~\ref{fig:dc}d overcomes the other torques originating from the magnetic anisotropies.
These behaviors are summarized in Fig.~\ref{fig:dc}g, where the reorientation time is plotted as a function of $F_0$ for different values of the biaxial anisotropy $K_{xy}$ and the Gilbert damping constant $\alpha$.
Overall, $t_\mathrm{r}$ is inversely proportional to $F_0$, which means that the vector potential $\sim F_0t_\mathrm{r}$ governs the reorientation dynamics; $t_\mathrm{r}$ becomes shorter as $K_{xy}$ or $\alpha$ decreases.
It is also found that the product $t_\mathrm{r}K_{xy}$ is a function of $F_0/K_{xy}$ for $K_{xy}\geq 0.01$ and $\alpha=1$.
The reorientation time shows the step-like feature indicated by the dotted lines, which reflects a discontinuous behavior of $\varphi(F_0 t)$ with respect to $F_0$ as seen in the inset of Fig.~\ref{fig:dc}f.
Although the anisotropy energies and the Gilbert damping constant are highly material dependent, this reorientation can occur in the timescale of subpicoseconds with a sufficiently large $F_0\ (\gtrsim 0.01)$ irrespective of $K_{xy}$ and $\alpha$.

\subsection*{Irradiation of a monocycle pulse}

\begin{figure*}[t]\centering
\includegraphics[width=1\hsize]{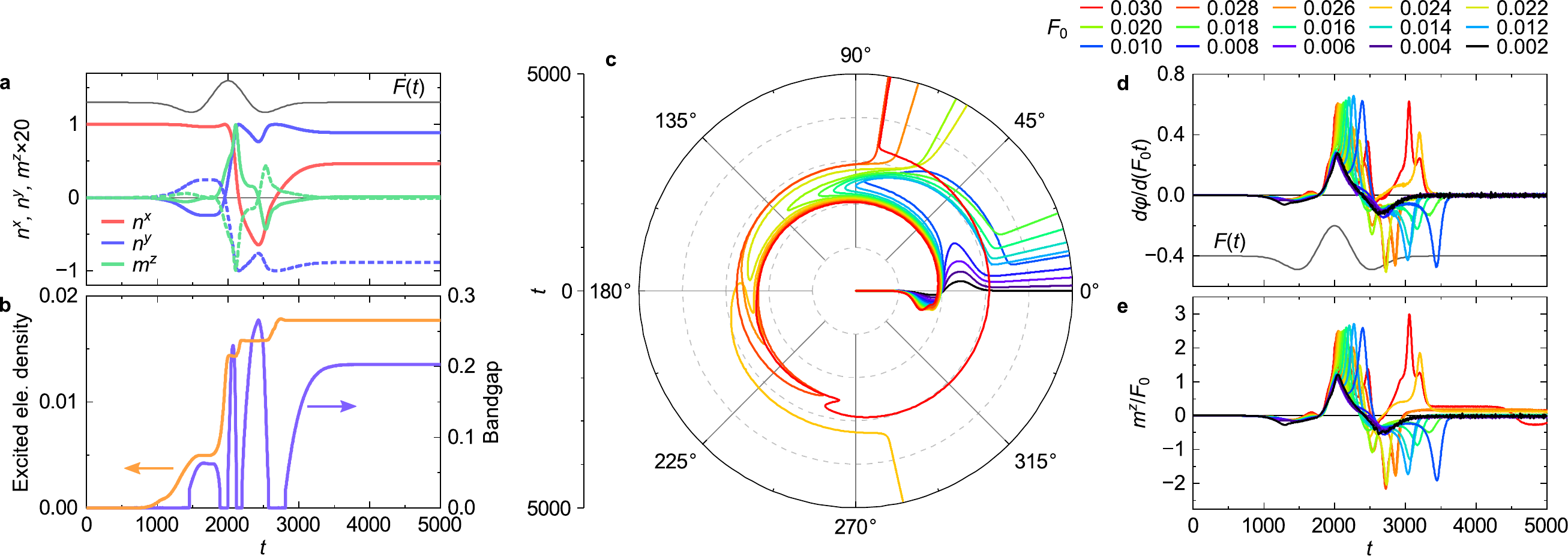}
\caption{Real-time dynamics induced by the monocycle pulse.
\textbf{a}, \textbf{b}~The time evolution of (\textbf{a}) the N\'eel vector and uniform magnetization, and (\textbf{b}) excited electron density and energy bandgap, as functions of time $t$, for $\bm{F}_0=F_0(\cos\pi/4,\sin\pi/4,0)$ and $F_0=0.02$.
The dashed lines in \textbf{a} denote the results for the $[\bar{1}10]$ polarization.
\textbf{c}~Polar plots of the in-plane angle of the N\'eel vector with respect to time $t$ for $F_0=0.002$--$0.03$.
\textbf{d},~\textbf{e}~The time derivative of the in-plane N\'eel vector angle and the $z$ component of the uniform magnetization divided by $F_0$ for $F_0=0.002$--$0.03$.
}
\label{fig:pulse}
\end{figure*}

We show the real-time dynamics induced by a monocycle pulse of which the frequency is in the terahertz range but is off-resonant (see Methods for parameters).
The electric field is parallel to the $[110]$ direction.
Figure~\ref{fig:pulse}a shows the in-plane components of $\bm{n}$ and the out-of-plane component of $\bm{m}$ for $F_0=0.02$.
The N\'eel vector rotates towards the $[010]$ direction during pulse irradiation, and it stops the rotation after irradiation.
This reorientation accompanies the modulation of the bandgap as shown in Fig.~\ref{fig:pulse}b.
When the bandgap is opened, the excited electron density does not increase as in the case of the constant field, which means that the pulse is off-resonant.

The dashed lines in Fig.~\ref{fig:pulse}a shows $n^y$ and $m^z$ for the same field amplitude but different polarization, i.e., $\bm{F}_0 \parallel [\bar{1}10]$, whose signs are opposite to those in the $\bm{F}_0 \parallel [110]$ case (solid lines).
This polarization dependence is also seen in the case of the constant field (not shown), and it implies that the direction of the N\'eel vector is determined by the polarization and inverted by the mirror operation on $\bm{F}_0$ against a plane perpendicular to $\bm{n}(t=0)$.

The pulse amplitude dependence of the N\'eel vector dynamics is summarized in Fig.~\ref{fig:pulse}c.
When the amplitude is weak ($F_0<0.01$), the N\'eel vector cannot climb over the potential barrier of the magnetic anisotropy.
However, the in-plane angle exceeds $45^\circ$ for a sufficiently large $F_0 \ (\geq 0.01)$ for which the excited electron density becomes greater than approximately $0.01$, and the N\'eel vector remains in the $[010]$ direction after irradiation for even larger $F_0\ (\geq 0.02)$.

Figures~\ref{fig:pulse}d and \ref{fig:pulse}e show the in-plane angular velocity of the N\'eel vector and the $z$ component of the uniform magnetization, both of which are divided by $F_0$.
For weak $F_0 < 0.01$, the time profiles of $d\varphi/d(F_0t)$ and $m^z/F_0$ are on each universal curve, whereas the deviations from the curves are found for $F_0 \geq 0.01$ and they lead to the reorientation towards the $[010]$ direction.
These time profiles are quite similar because $d\varphi/dt$ is approximated by $2K_zm^z$ as long as $\Vert \bm{m} \Vert \ll \Vert \bm{n} \Vert$.

\subsection*{Optical conductivity and magneto-optical effect}

\begin{figure*}[t]\centering
\includegraphics[width=0.92\hsize]{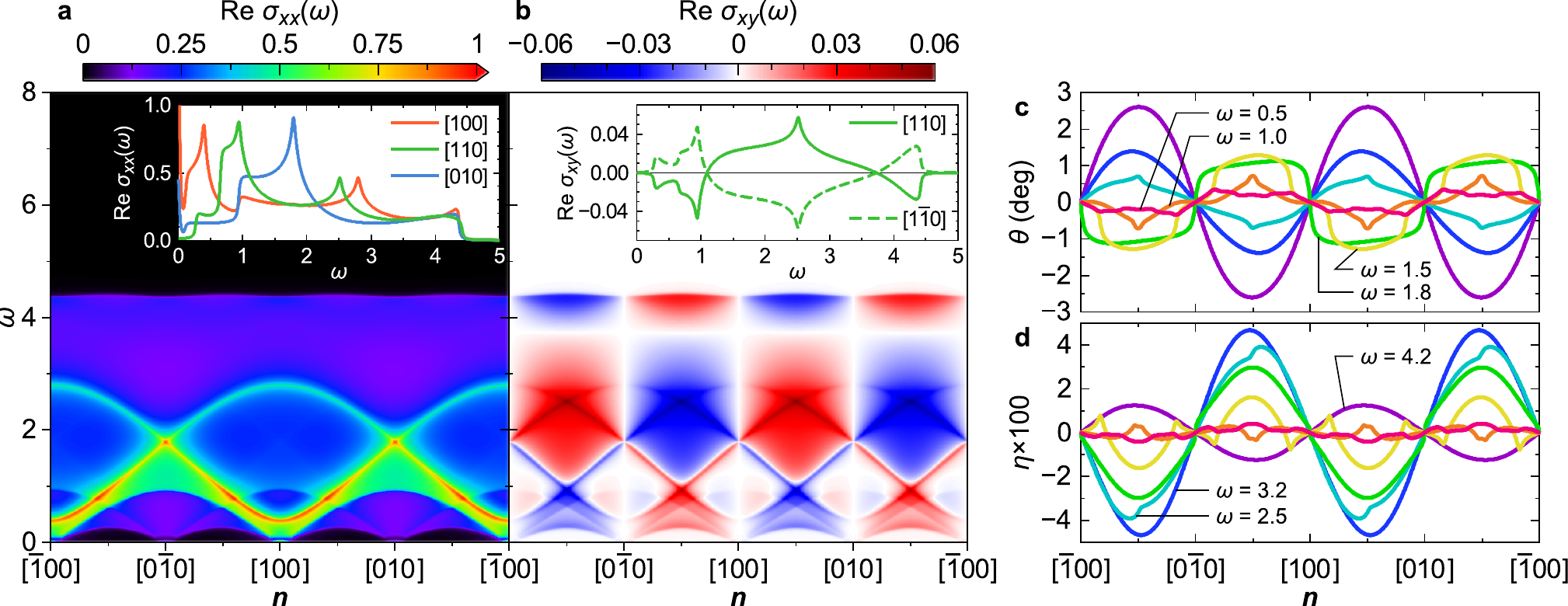}
\caption{Optical conductivity and magneto-optical effect.
\textbf{a}, \textbf{b}~Real parts of (\textbf{a})~longitudinal and (\textbf{b})~transverse optical conductivities.
Energy $\omega$ and optical conducvitity $\sigma$ are shown in units of $h_1 = 1\ \mathrm{eV}$ and $e^2/(\hbar a) = 6.4\times 10^3\ \mathrm{S\, cm^{-1}}$, respectively.
Insets in (\textbf{a}) and (\textbf{b}) show the optical conductivity for $\bm{n} \parallel [100], [110], [010]$, and $[1\bar{1}0]$.
Panels \textbf{c}, \textbf{d} show the magneto-optical rotation angle $\theta$ and ellipticity $\eta$ as functions of the in-plane angle of the N\'eel vector.
The lattice constant and broadening factor are set to $a = 0.38\ \mathrm{nm}$ and $\varGamma/h_1=0.02$, respectively.
}
\label{fig:conductivity}
\end{figure*}

Finally, we show the optical longitudinal and transverse (Hall) conductivities for different N\'eel vector angles, and we discuss the resulting magneto-optical Voigt effect, assuming that the system is in the collinear AFM ground states.
Figures~\ref{fig:conductivity}a and \ref{fig:conductivity}b display the real parts of the longitudinal ($\sigma_{xx}$) and transverse ($\sigma_{xy}$) conductivities as functions of the in-plane N\'eel vector angle $\varphi$.
The peaks of $\real \sigma_{xx}(\omega)$ for $\omega>0$ are attributed to the Van Hove singularity; the Drude peak is also seen at $\omega = 0$ for $\bm{n}\parallel [100]$ and $[010]$ when $h_2 \neq 0$.
Because of the two-fold symmetry around the $z$ axis, $\sigma_{xx}$ and $\sigma_{xy}$ are periodic in $\varphi$.
Here, $\sigma_{yy}$ is equivalent to $\sigma_{xx}$ under the rotation $\varphi \mapsto \varphi+\pi/2$, and $\sigma_{xy}$ is equal to $\sigma_{yx}$.
The sign of the transverse conductivity is unchanged by $\lambda\mapsto -\lambda$ or $J_\mathrm{exc} \mapsto -J_\mathrm{exc}$ (not shown), which implies the conductivity is an even function of $\lambda$ and $J_\mathrm{exc}$.
Since the present model holds the $\mathrm{D_{4h}}$ symmetry in the limit of $J_\mathrm{exc}\bm{n}\to 0$, the conductivity linear in the electric field can be expanded with respect to a time-reversal-breaking field $\bm{B}\sim J_\mathrm{exc}\bm{n}$ as $\sigma_{ij}(\bm{B}) = \sigma_{ij}^{(0)} + \sigma_{ijk}^{(1)}B_k + \sigma_{ijkl}^{(2)}B_kB_l$, where $\sigma^{(1)}$ vanishes when $\bm{B}\perp [001]$ and
\begin{align}
\sigma^{(2)}_{ij}=
\begin{bmatrix}
\sigma^{(2)}_{xxxx}B_x^2+\sigma^{(2)}_{xxyy}B_y^2 & 2\sigma^{(2)}_{xyxy}B_xB_y \\
2\sigma^{(2)}_{xyxy}B_xB_y & \sigma^{(2)}_{xxyy}B_x^2+\sigma^{(2)}_{xxxx}B_y^2 \\
\end{bmatrix}_{ij}
\end{align}
is a symmetric tensor.
The off-diagonal components of $\sigma(\bm{B})$ can be written as $\sigma_{xy}(\bm{B})=\sigma_{yx}(\bm{B})\propto \sin 2\varphi$ for $\bm{n}=(\cos\varphi,\sin\varphi,0)$.
This is consistent with the numerical result of $\sigma_{xy}$, although the present values of $\lambda\ (=0.8)$ and $J_\mathrm{exc}\ (=0.6)$ are beyond the perturbative regime.
From these properties of $\sigma_{xx}$ and $\sigma_{xy}$, we can identify the direction in which the N\'eel vector rotates by measuring the longitudinal optical conductivity and magneto-optical signal.
Figures~\ref{fig:conductivity}c and \ref{fig:conductivity}d show the magneto-optical rotation angle and ellipticity calculated from $\sigma_{xx}$ and $\sigma_{xy}$ (see Methods).
The rotation angle $\theta$ exhibits large values up to $1^\circ$.
In real materials, the values of $\lambda$ and $J_\mathrm{exc}$ may be smaller than the values adopted here, which causes a reduction in the rotation angle down to a few millidegrees.
Even so, the magneto-optical rotation can be measured using modern optical apparatus.

The transverse conductivity $\sigma_{xy}$ appears when $\lambda\neq 0$ and $J_\mathrm{exc}\neq 0$ even though the magnetic structure is collinear AFM.
Recently, the transverse response in collinear antiferromagnets is attracting considerable attention, and it has been discussed for the case in which the conductivity tensor is antisymmetric~\cite{Smejkal2020,Naka2020}; our calculations demonstrate the transverse optical response attributed to the symmetric tensor in collinear antiferromagnets.

\section*{Discussions} \label{sec:discussion}
We investigated the real-time dynamics of the magnetic and electronic structures induced by the constant and pulse electric fields, and we revealed the microscopic mechanism of the NSOT-driven reorientation of the N\'eel vector.
Our method can be applied to any tight-binding model coupled to localized moments.
The experimental verification of the present results is necessary for a better understanding of the NSOT-induced dynamics and for further progress in ultrafast AFM spintronics combined with the topological aspects of the materials.
Candidate materials include orthorhombic CuMnAs~\cite{Tang2016,Smejkal2017,Kim2018,Xu2020} and MnPd$_{2}$~\cite{Shao2019}, which have been studied in the field of AFM spintronics and are expected to host Dirac fermions.
Note that the reorientation of the N\'eel vector demonstrated in this study is not directly related to what has been observed in experiments because of the discrepancy in the timescale.
To make more quantitative discussions on, e.g., the timescale of the reorientation and the required magnitude of the field, we should consider the crystal structure and dimensionality of materials, obtain realistic estimates of model parameters as well as the Gilbert damping constant from the ab initio calculations and experiments, and take other dissipation processes caused by the electron-electron interaction into account; these are left for future work.

Recent intense terahertz pulse sources, whose oscillation period is $1\ \mathrm{ps}$ and peak amplitude is more than $1\ \mathrm{MV\, cm^{-1}}$~\cite{Fulop2020}, are promising to drive the reorientation of the N\'eel vector in the picosecond timescale.
In our calculations, a typical value of $F_0$ is of order $0.01$, which corresponds to $0.26\ \mathrm{MV\, cm^{-1}}$ (see Methods).
There are many earlier studies on the electrical modulation and detection of the N\'eel vector~\cite{Amin2020}; their time resolution was limited to $\gtrsim 1\ \mathrm{ns}$.
However, it may be possible to observe the N\'eel vector dynamics in real time via optical measurements because the magnetic structure is closely related to the electronic structure in the present model.
One approach is the time- and angle-resolved photoemission spectroscopy; this can directly access the energy band structure, which reflects the direction of the N\'eel vector as the positions and energy gaps of the Dirac points or nodal lines.
Another approach is to use magneto-optical effects.
This system exhibits the magneto-optical rotation with angles up to $1^\circ$.
The time-resolved magneto-optical measurements can be carried out even in the subpicosecond timescale, which provides a sufficient time resolution to observe the real-time dynamics discussed in this study.

\appendix
\section*{Methods} \label{sec:methods}

\subsection*{Real-time dynamics}
We calculate the real-time dynamics of the itinerant electrons and the localized magnetic moments as follows~\cite{Chern2018,Koshibae2009}.
The quantum state of the electrons is exactly described by a one-body density matrix $\hat{\rho}_{\bm{k}}$ because the electronic Hamiltonian $\mathcal{H}_\mathrm{ele}$ contains no many-body interaction.
Once the localized magnetic moments are given, the Hamiltonians in Eqs.~\eqref{eq:hamiltonian_ele} and \eqref{eq:hamiltonian_exc} can be diagonalized for each $\bm{k}$ as $(\mathcal{H}_\mathrm{ele}+\mathcal{H}_\mathrm{exc})\vert\bm{k}\nu\rangle = \varepsilon_{\bm{k}\nu}\vert\bm{k}\nu\rangle$, where $\vert\bm{k}\nu\rangle$ is obtained by the unitary transform as $\vert\bm{k}\nu\rangle = \sum_{\gamma = \mathrm{A},\mathrm{B}} \sum_{s ={\uparrow},{\downarrow}} \vert \bm{k}\gamma s\rangle U_{\nu,\gamma s}^*$.
In the initial state in which the external field is absent, the density matrix is given by $\tilde{\rho}_{\bm{k}} = \sum_{\nu} \vert\bm{k}\nu\rangle n_{\bm{k}\nu} \langle \bm{k}\nu \vert$, where $n_{\bm{k}\nu}=\varTheta (\mu-\epsilon_{\bm{k}\nu})$ is the step function with $\mu$ being the chemical potential chosen such that the number of electrons is $N_\mathrm{ele}$.
The time evolution of the density matrix is governed by the von Neumann equation
\begin{align}
\frac{d\hat{\rho}_{\bm{k}}}{dt} = -\mathrm{i} [\mathcal{H},\hat{\rho}_{\bm{k}}] \label{eq:neumann}
\end{align}
with the initial condition $\hat{\rho}_{\bm{k}}(t=0) = U_{\bm{k}}^{-1}\tilde{\rho}_{\bm{k}}U_{\bm{k}}$.
The expectation value is defined by $\langle \cdot \rangle = N^{-1} \sum_{\bm{k}} \Tr [\,\cdot\, \hat{\rho}_{\bm{k}}]$, where $N$ represents the number of $\bm{k}$ points in the Brillouin zone.
The localized moments on each sublattice denoted by $\bm{S}_\mathrm{A}$ and $\bm{S}_\mathrm{B}$ are treated as classical unit vectors, and therefore, the Hamiltonian $\mathcal{H}(t)$ depends on the classical spins and the external field.
The classical spins follow the LLG equation
\begin{align}
\frac{d\bm{S}_\gamma}{dt} = \bm{S}_\gamma \times \bm{b}_\gamma - \alpha \bm{S}_\gamma \times \frac{d\bm{S}_\gamma}{dt} \label{eq:llg}
\end{align}
for $\gamma=\mathrm{A},\mathrm{B}$, where $\alpha$ denotes the Gilbert damping constant and $\bm{b}_{\gamma} \equiv -\langle\partial\mathcal{H}/\partial\bm{S}_\gamma\rangle$ represents the effective field given by
\begin{align}
\bm{b}_\gamma &= -J_\mathrm{exc} \langle \bm{\sigma}_\gamma \rangle
-2{\left( K_{xy}S_\gamma^x (S_\gamma^y)^2, K_{xy}(S_\gamma^x)^2 S_\gamma^y,K_z S_\gamma^z \right)} .
\end{align}
Here, $\bm{\sigma}_{\mathrm{A}/\mathrm{B}} = (1\pm\tau^z)\bm{\sigma} /2$ denotes the sublattice spin density.
The time-dependent electric field $\bm{F}(t)$ is incorporated in the model via the Peierls substitution $\mathcal{H}_\mathrm{ele}(\bm{k}) \mapsto \mathcal{H}_\mathrm{ele}(\bm{k}-e\bm{A}(t)a/\hbar)$, where $\bm{A}(t) = -\int^t\bm{F}(t')\,dt'$, $e$, and $a$ denote the vector potential, elementary charge, and lattice constant, respectively.
Throughout this paper, $h_1$, $e$, $a$, and $\hbar$ are set to unity.
The physical quantities of time, electric-field magnitude, and conductivity are expressed in units of $\hbar/h_1=0.66\ \mathrm{fs}$, $h_1/(ea) = 26.3\ \mathrm{MV\, cm^{-1}}$, and $e^2/(\hbar a) = 6.4\times 10^3\ \mathrm{S\, cm^{-1}}$, respectively, for $h_1=1\ \mathrm{eV}$ and $a=0.38\ \mathrm{nm}$.
In most calculations, we use $h_2=0.08$, $\lambda=0.8$, $J_\mathrm{exc}=0.6$, $K_z=0.1$, $K_{xy}=0.01$, and $\alpha=1$.

The easy-plane anisotropy coefficient $K_z$ determines the in-plane angular velocity of the N\'eel vector $d\varphi/dt$.
For the above parameters, the difference in the total energies of the ground states is $\langle\mathcal{H}\rangle\vert_{\bm{n}=(0,0,1)} - \langle\mathcal{H}\rangle\vert_{\bm{n}=(1,0,0)} = 0.13h_1$.
This may be considerably larger than the anisotropy energy of real materials ($\lesssim 1\ \mathrm{meV}$)~\cite{Kim2018,Xu2020}.
However, we confirmed that the reorientation of the N\'eel vector occurs within $10\ \mathrm{ps}$ even for smaller values: $K_z=K_{xy}=0.001$, $\lambda=0.08$, $J_\mathrm{exc}=0.06$, and $F_0=0.001$.

Equations~\eqref{eq:neumann} and \eqref{eq:llg} are simultaneously solved using the fourth-order Runge--Kutta method with a time step $\delta t=0.01$.
We perform the numerical calculation with $N=512 \times 512$-point mesh in the Brillouin zone, and we fix the number of the electrons to $N_\mathrm{ele}=2N$ (half-filled).
The external field $\bm{F}(t) = -\partial_t \bm{A}(t)$ is given by $\bm{A}(t) = -\bm{F}_0 t \varTheta(t)$ for the constant electric field and $\bm{A}(t) = -\bm{F}_0 {\cdot} (t-t_\mathrm{c}) \exp[-(t-t_\mathrm{c})^2/(2t_\mathrm{w}^2)]$ for the monocycle pulse.
We adopt $t_\mathrm{c}=2000$ and $t_\mathrm{w}=300$.

\subsection*{Optical conductivity and magneto-optical effect}
We calculate the optical conductivity from
\begin{align}
\sigma_{ij}(\omega) &= \frac{\chi_{ij}(\omega+\mathrm{i}\varGamma)-\chi_{ij}(\mathrm{i}\varGamma)}{\mathrm{i}(\omega+\mathrm{i}\varGamma)}, \\
\chi_{ij}(z) &= \frac{1}{N} \sum_{\bm{k}\mu\nu} J_{\bm{k}\mu\nu}^{i} J_{\bm{k}\nu\mu}^{j} \frac{n_{\bm{k}\mu}-n_{\bm{k}\nu}}{\varepsilon_{\bm{k}\nu}-\varepsilon_{\bm{k}\mu}-z},
\end{align}
where $\bm{J}_{\bm{k}\mu\nu}=\langle\bm{k}\mu\vert \bm{J}_{\bm{k}} \vert\bm{k}\nu\rangle$ denotes the matrix element of the electric current operator $\bm{J}_{\bm{k}}=-\delta\mathcal{H}(\bm{k}-\bm{A})/\delta\bm{A} \vert_{\bm{A}=0}$ and $\varGamma$ represents a broadening factor.
The dielectric tensor is given by
\begin{align}
\epsilon_{ij}(\omega) = \delta_{ij} + \frac{\mathrm{i}\sigma_{ij}(\omega)}{\epsilon_{0}\omega}
\end{align}
with $\epsilon_0=8.854\ \mathrm{pF\, m^{-1}}$ being the electric constant.
Since the optical conductivity tensor is symmetric ($\sigma_{xy}=\sigma_{yx}$) and anisotropic ($\sigma_{xx}\neq \sigma_{yy}$) in the present model, we cannot use the well-known formula of the complex Kerr rotation angle $\varPhi_\mathrm{K} = \epsilon_{xy}/[(1-\epsilon_{xx})\sqrt{\epsilon_{xx}}]$.
According to Maxwell's equations in material media, the electric-field amplitude $\bm{F}_0$ satisfies
\begin{align}
(\hat{\bm{N}}\cdot\hat{\bm{N}})\bm{F}_0 - (\hat{\bm{N}}\cdot\bm{F}_0)\hat{\bm{N}} + \epsilon\bm{F}_0 = 0, \label{eq:refractive}
\end{align}
where $\hat{\bm{N}}$ represents a complex refractive index vector.
From Eq.~\eqref{eq:refractive} we have the relation $(\hat{N}^2-\epsilon_{xx})(\hat{N}^2-\epsilon_{yy})=\epsilon_{xy}^2$, assuming that the propagation vector of incident light is perpendicular to the two-dimensional $xy$-plane.
The symmetric property of the conductivity and dielectric tensors leads to the eigenmodes of Eq.~\eqref{eq:refractive} being linearly polarized, contrary to the antisymmetric case in which the eigenmodes are circularly polarized.
The boundary condition is that the tangential components of the electric field $\bm{F}$ and magnetic field $\hat{\bm{N}}\times\bm{F}/c_0$ ($c_0$ is the speed of light) are continuous at the surface, from which we obtain two eigenmodes and complex refractive indices.
The incident linearly polarized light is decomposed into a linear combination of the eigenmodes.
The amplitude vector of the reflected light is proportional to $(r_\mathrm{pp},r_\mathrm{sp},0)$, where $r_\mathrm{pp}$ and $r_\mathrm{sp}$ are the amplitude reflection coefficients of the incident p-polarized light to the reflected p- and s-polarized light.
The magneto-optical rotation angle $\theta$ and ellipticity $\eta$ are determined by the fit of the electric-field trajectory to an ellipse rotated around the $z$ axis by angle $\theta$, and the relation $\eta=\sgn[\arg(r_\mathrm{sp}/r_\mathrm{pp})] \times \vert r_\mathrm{sp}\vert/\vert r_\mathrm{pp}\vert$, respectively.

\begin{acknowledgments}
One of the authors, Sumio~Ishihara, passed away on 7 November 2020 during the preparation of the manuscript.
We thank H.~Matsueda and Y.~Masaki for their valuable comments and critical reading of the manuscript.
This work was supported by JSPS KAKENHI Grant Nos.~JP19K23419, JP20K14394, JP17H02916, JP18H05208, and JP20H00121.
The numerical calculations were performed using the facilities of the Supercomputer Center, the Institute for Solid State Physics, the University of Tokyo.
\end{acknowledgments}

\section*{Author contributions}
A.O. conceived the idea, performed the calculations, and wrote the manuscript.
A.O. and S.I. discussed and interpreted the results.

\bibliography{reference}

\end{document}